# Pediatric lymphoma may develop by "one-step" cell transformation of a lymphoid cell

Jicun Wang-Michelitsch[1]*, Thomas M Michelitsch[2]

[1] Independent researcher,

[2] Sorbonne Université, Institut Jean le Rond d'Alembert, CNRS UMR 7190 Paris, France

## Abstract

Lymphomas are a large group of neoplasms developed from lymphoid cells (LCs) in lymph nodes (LNs) or lymphoid tissues (LTs). Some forms of lymphomas, including Burkitt lymphoma (BL), $ALK^+$ anaplastic large cell lymphoma ($ALK^+$-ALCL), and T-cell lymphoblastic lymphoma/leukemia (T-LBL), occur mainly in children and teenagers. Hodgkin's lymphoma (HL) has a peak incidence at age 20s. To understand pediatric lymphoma, we have recently proposed two hypotheses on the causes and the mechanism of cell transformation of a LC. Hypothesis **A** is**:** repeated bone-remodeling during bone-growth and bone-repair may be a source of cell injuries of marrow cells including hematopoietic stem cells (HSCs), myeloid cells, and LCs; and thymic involution may be a source of damage to the developing T-cells in thymus. Hypothesis **B** is: a LC may have three pathways on transformation: a slow, a rapid, and an accelerated. In this paper, we discuss pediatric lymphomas by this hypothesis. Having a peak incidence at young age, BL, T-LBL, $ALK^+$-ALCL, and HL develop more likely as a result of transformation of a LC via rapid pathway. In BL, $ALK^+$-ALCL, and HL, the cell transformations may be triggered by severe viral infections. In T-LBL, the cell transformation may be related to thymic involution. Occurring in both adults and children, diffuse large B-cell lymphoma (DLBCL) may develop via slow or accelerated pathway. In conclusion, pediatric lymphoma may develop as a result of "one-step" cell transformation of a LC; and severe viral infections may be the main trigger for the rapid transformation of a LC in a LN/LT.

**This paper has the following structure:**

**I. Introduction**

**II. Different forms of lymphomas occur at different ages**

  2.1 Forms of lymphomas occurring mainly in children and teenagers
      2.1.1 Burkitt lymphoma (BL)
      2.1.2 $ALK^+$-anaplastic large cell lymphoma ($ALK^+$-ALCL)
      2.1.3 T-lymphoblastic lymphoma/leukemia (T-LBL)
  2.2 Forms of lymphomas occurring in both adults and children
      2.2.1 Hodgkin's lymphoma (HL)
      2.2.2 Diffuse large B-cell lymphoma (DLBCL)
  2.3 Forms of lymphomas occurring mainly in adults
      2.3.1 Follicular lymphoma (FL)
      2.3.2 Mucosa-associated lymphoid tissue lymphoma (MALTL)
      2.3.3 Mantle cell lymphoma (MCL)
      2.3.4 Adult T-cell lymphoma/leukemia (ATLL)

**III. Three potential sources of cell injuries of lymphoid cells (LCs)**

  3.1 Repeated bone-remodeling during bone-growth and bone-repair in marrow cavity
  3.2 Long-term thymic involution in thymus
  3.3 Pathogen-infections in lymph nodes (LNs) and lymphoid tissues (LNs)

**IV. DNA changes are generated and accumulate in cells as a consequence of repeated cell injuries**

**V. A LC may have three pathways on cell transformation**

**VI. The age of occurrence of lymphoma is determined by the transforming pathway of a LC**

  6.1 A lymphoma occurring mainly in adults: via slow pathway
  6.2 A lymphoma occurring at any age without increasing incidence with age: via rapid pathway
  6.3 A lymphoma occurring at any age with increasing incidence with age: via accelerated pathway

**VII. Pediatric lymphoma may develop as a result of rapid cell transformation of a LC**

  7.1 BL: as a result of rapid transformation of a centroblast by severe viral infections
  7.2 $ALK^+$-ALCL: is it a result of rapid cell transformation of a "T-immunoblast"?
  7.3 T-LBL: is it a result of rapid transformation of a T-lymphoblast by thymic regression?
  7.4 HL: is it a result of rapid transformation of a B-immunoblast by complex karyotype?
  7.5 DLBCL: as a result of transformation of a B-immunoblast/plasmablast via slow or accelerated pathway
  7.6 FL, MCL, MATLTL, and ATLL: related to not only the cell injuries of LCs in LNs/LTs but also that of hematopoietic stem cells (HSCs) and LCs in marrow

**VIII. Conclusions**

## I. Introduction

Lymphomas are a large group of neoplasms developed from lymphoid cells (LCs). Lymphomas occur not only in adults but also in children. Some forms of lymphomas, including Burkitt lymphoma (BL), ALK+ anaplastic large cell lymphoma (ALK$^+$-ALCL), and T-lymphoblastic lymphoma/leukemia (T-LBL), have a peak incidence in children or adolescents. These pediatric lymphomas are the most aggressive forms of lymphomas. Hodgkin's lymphoma (HL) has a peak incidence in young adults in western countries. Diffuse large B-cell lymphoma (DLBCL) occurs mainly in adults, but also in children. Differently, some forms of lymphomas occur mainly in old people, and they include follicular lymphoma (FL) and mucosa-associated lymphoid tissue lymphoma (MALTL) (Figure 1).

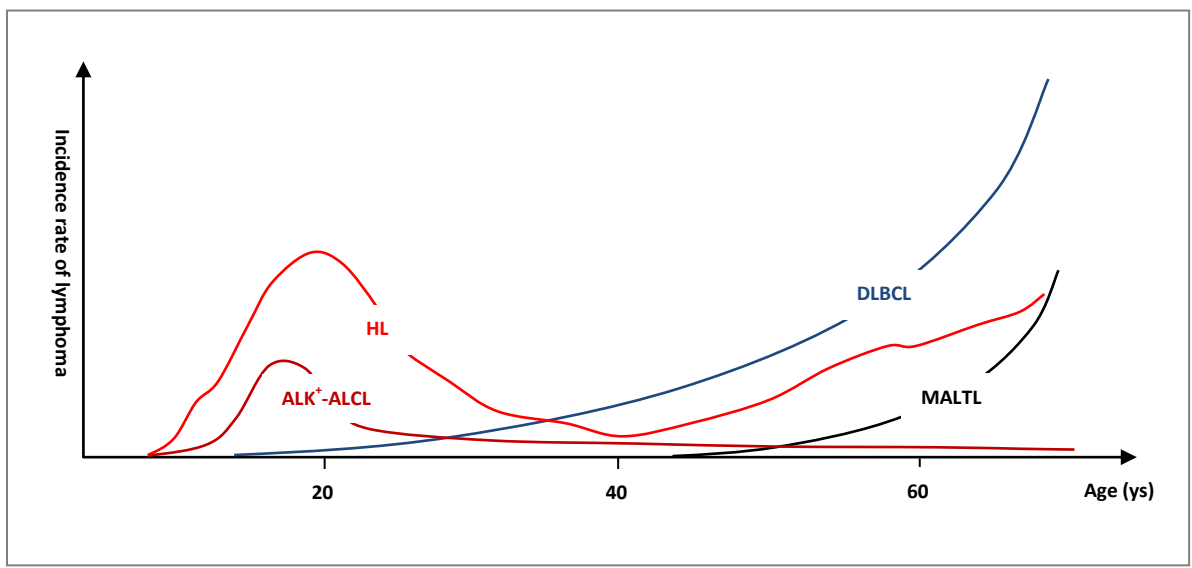

**Figure 1. Age-specific incidences of lymphomas (a schematic graph)**

Lymphoma occurs not only in adults but also in children. Different forms of lymphomas tend to occur at different ages. For example, ALK+ anaplastic large cell lymphoma (ALK$^+$-ALCL) has a peak incidence at age 15-20 (**dark red line**). Hodgkin's lymphoma (HL) has a peak incidence at age 15-25 (**red line**). Diffuse large B-cell lymphoma (DLBCL) occurs in adults and children, and has increasing incidence with age (**blue line**). Mucosa-associated lymphoid tissue lymphoma (MALTL) occurs mainly in old people and has increasing incidence with age (**black line**).

Different forms of lymphomas arise often from LCs at different developing stages. "LCs" include all the cells of lymphoid linage in bone marrow, thymus, lymph nodes (LNs), and lymphoid tissues (LTs). DNA changes including gene mutations and chromosome changes (CCs) are the drivers for cell transformation of a LC. However, different forms of DNA changes may drive cell transformation in different manners. To understand pediatric lymphomas and lymphoid leukemias, we proposed recently a hypothesis: a LC may have three pathways on cell transformation: a slow, a rapid, and an accelerated (Wang-Michelitsch, 2018b). In the present paper, we will discuss the developing mechanism of pediatric lymphomas by this hypothesis in comparison with adult lymphomas. We aim to show by our discussion that, pediatric lymphoma may develop as a result of one-step cell transformation of a LC by a great-effect chromosome change.

We use the following abbreviations in this paper:

| | |
|---|---|
| ABC: activated B-cell like DLBCL | HIV: human immunodeficiency virus |
| AIDS: acquired immunodeficiency syndrome | HL: Hodgkin's lymphoma |
| ATLL: adult T-cell lymphoma/leukemia | HLTV-1: human T cell lymphotropic virus type 1 |
| ALK$^+$-ALCL: ALK-positive anaplastic large cell lymphoma | HSC: hematopoietic stem cell |
| | IECC: intermediate-effect chromosomal change |
| ALL: acute lymphoblastic leukemia | LN: lymph node |
| BL: Burkitt lymphoma | LT: lymphoid tissue |
| CC: chromosome change | MALTL: mucosa-associated lymphoid tissue lymphoma |
| DHL: double hit lymphoma | |
| DLBCL: diffuse large B-cell lymphoma | MCL: mantle cell lymphoma |
| EBV: Epstein-Barr virus | MECC: mild-effect chromosomal change |
| FL: follicular lymphoma | NHL: non-Hodgkin's lymphoma |
| GCB: germinal center B-cell like DLBCL | PDM: point DNA mutation |
| GECC: great-effect chromosomal change | T-LBL: T- lymphoblastic lymphoma/leukemia |

## II. Different forms of lymphomas occur at different ages

There are two groups of lymphomas: Hodgkin's lymphomas (HLs) and non-Hodgkin's lymphomas (NHLs). HLs are a group of lymphomas characterized by giant Reed-Sternberg (R-S) cells in tumor tissues. The lymphomas that do not have R-S cells belong to the group of NHLs. NHLs account for 90% of total lymphomas. Over 60 forms of NHLs have been identified (Swerdlow, 2016). Most forms of lymphomas have higher incidences in male than in female. In this part, we make firstly a brief review on the major forms of lymphomas on their developing age, pathology, cell of origin, and recurrent DNA changes in tumor cells.

### 2.1 Forms of lymphomas occurring mainly in children and teenagers

Three forms of NHLs, including BL, T-LBL, and ALK$^+$-ALCL, occur mainly in children and adolescents (Sandlund, 2015). BL occurs more often in children younger than age 10, whereas ALK$^+$-ALCL and T-LBL occur mainly at age 15-20 (Table 1). BLs, T-LBLs, and ALK$^+$-ALCLs represent respectively 30%-40%, 20%-30%, and 10%-15% of pediatric NHLs (< age 15) (Chung, 2016). These pediatric forms of NHLs are different to each other on cell of origin and on causing factor, but they are all aggressive forms of lymphomas (Ansell, 2015a and 2015 b).

### 2.1.1 Burkitt lymphoma (BL)

BL is the most aggressive form of lymphoma. BL has three subtypes: endemic (African), sporadic (non-African), and immunodeficiency-associated. Endemic BL occurs mainly in children in sub-Saharan Africa, and the cases account for 25% of total BLs (Rochford, 2015). Endemic BL has a peak incidence at age 4-7, and 30%-40% of endemic BLs occur before age 5 (Table 1). Endemic BL starts often from the nasopharyngeal mucosa of the patient and develops rapidly into mandible, jaw, the skull, and center nervous system. Endemic BL rarely affects superficial LNs and does not develop into leukemia. Immunodeficiency-associated BL is a subtype of BL that develops as a consequence of infection of human immunodeficiency virus (HIV). A patient with acquired immunodeficiency syndrome (AIDS) caused by HIV-infection may have 1000 times of risk higher than a healthy people on BL development. Sporadic BL is the BL that occurs outside of Africa and is not related to HIV-infection. Sporadic BL occurs at all ages but more often in children and young adults. Sporadic BLs represent 40% of BLs in western countries. Sporadic BL and immunodeficiency-associated BL start often from an organ in abdomen and soon affect bone marrow.

**Table 1. Pediatric forms of lymphomas**

| Form of lymphoma | Incidence rate (< age 15) | Age of peak incidence | Percentage of cases of age <5 in total cases of age <15 |
| --- | --- | --- | --- |
| BL | 30%-40% of pediatric NHLs | Age 4-7 | 30%-40% |
| ALK$^+$-ALCL | 10%-15% of pediatric NHLs | Age 15-20 | 5%-15% |
| T-LBL | 20%-30% of pediatric NHLs | Age 15-20 | 20%-30% |
| HL | 20%-30% of pediatric lymphomas | Age 15-25 | Rare |
| DLBCL | 8%-10% of pediatric NHLs | No peak | Rare |

Endemic BL may be related to the co-infection of Epstein-Barr virus (EBV) and malarial in children in Africa. It is thought that EBV may be the direct cause for cell transformation of a LC and malaria may play multiple and simultaneous roles in BL etiology (Moormann, 2016). Differently, only 20% of sporadic BLs are related to EBV-infection. All subtypes of BL arise from a small non-cleaved B-cell, probably a centroblast, in germinal center of a LN (Table 2). The most frequent DNA changes in BL are translocations of Ig/MYC, including t(8;14) (IgH/MYC), t(2;8) (IgK/MYC), and t(8;22) (IgL/MYC) (Nguyen, 2017). t(8;14) is found in 70%-80% of BLs, and each of t(2;8) and t(8;22) is found in 5%-10% of BLs. These forms of translocations result in generation of fusion gene of *Ig-MYC* and constant over-expression of *MYC*. Thus, translocations of Ig/MYC are the main driver DNA changes in BL development. BL is sensitive to chemotherapy, and complete remission of BL by standard treatment is 90%. More than 50% of pediatric BLs can be cured.

### 2.1.2 ALK$^+$-anaplastic large cell lymphoma (ALK$^+$-ALCL)

ALCL has three subtypes: ALK-positive (ALK$^+$) systemic, ALK-negative (ALK$^-$) systemic, and cutaneous. ALK, namely the anaplastic lymphoma kinase, is an abnormal form of a cell surface protein which has kinase activity. ALK$^+$-ALCL is the most common subtype of ALCL, and ALK$^+$-ALCLs represent 60% of ALCLs (Lowe, 2013). ALK$^+$-ALCL occurs mainly before age 30 and has a peak incidence at age 15-20 (Table 1). About 5%-15% of pediatric ALK$^+$-ALCLs occur before age 5. Differently, ALK-negative ALCL occurs in both adults and children, but more in adults. Cutaneous ALCL occurs mainly in old people. In an ALK$^+$-ALCL patient, the tumor starts often from the superficial LNs in neck, armpit, or groin. In 60% of ALK$^+$-ALCLs, the neoplasm has affected extra-nodal LTs in skin or abdomen at diagnosis (Table 2). In 30% of cases, bones are affected and bone pain can be the first symptom in these patients.

**Table 2. Age distributions and the cells of origin of major forms of lymphomas**

| Form of lymphoma | Age of diagnosis | Rate in NHLs | Cell of origin | Recurrent DNA changes | Starting locations |
|---|---|---|---|---|---|
| BL | Any age, peak at age 4-7 | 2% | Centroblast | t(8;14), t(2;8), or t(8;22) | Extra-nodal LTs in nasopharyngeal, mandible, and abdomen |
| ALK$^+$-ALCL | Any age, peak at age 15-20 | Rare | Activated T-cell | t(2; 5) | LNs and extra-nodal LTs in skin and abdomen |
| T-LBL | Any age, peak at age 15-20 | 2% | T-lymphoblast or pro-lymphocyte | t(7;9), t(1;14), or t(10;14) | Mediastinum |
| HL | Any age, peak at age 20s | 10% of total lymphomas | B-immunoblast? | Complex karyotypes | A LN in neck |
| DLBCL | Any age, Mostly > age 60 | 35% | B-immunoblast or plasmablast | MYC translocation + gene mutations | LNs in neck and extra-nodal LTs in abdomen |
| FL | > age 55 | 25 % | Centrocyte or centroblast | t(14;18) + gene mutations | A LN in neck |
| MCL | >age 55 | 6% | Naïve B-cell | t(11;14) and *SOX11* mutation | Spleen, multiple LNs, and extra-nodal LTs |
| MALTL | > age 55 | 7.5% | Memory B-cell | IRTA1, t(11;18), t(14;18), and gene mutations | Extra-nodal LTs in stomach |
| ATLL | > age 40 | Only in some countries | Memory or effector Th-cell | Gene mutations | Extra-nodal LTs in skin |

In pathology, ALK$^+$-ALCL is characterized by cohesive proliferation of large multi-morphologic blasts which invade into the sinus of affected LNs. The blasts are CD30-positive. CD30 is a type of cell cytokine receptor expressed in tumor cells. CD30 is negative in normal T-cells but positive in antigen-activated T-cells. CD30 is found also in R-S cells in HL. In 90% of ALK$^+$-ALCLs, the cell of origin is an activated T/NK-cell. 80% of ALK$^+$-ALCLs have t(2;5) in tumor cells. By generating a fusion gene of *NMP-ALK* and permanent activation of ALK, t(2; 5) is the main driver DNA change in ALK$^+$-ALCL development. In some ALK$^+$-ALCLs, HIV-infection may be a causing factor.

*2.1.3 T-lymphoblastic lymphoma/leukemia (T-LBL)*

Lymphoblastic lymphoma/leukemia (LBL) is also a highly aggressive form of lymphoma. LBLs represent 2% of NHLs (Table 2). 90% of LBLs are T-cell LBL (T-LBL), thus we discuss mainly T-LBL. T-LBL occurs before age 35, but mostly in teenagers. The peak incidence of T-LBL is at age 15-20. The T-LBLs occurred before age 5 account for 20%-30% of pediatric T-LBLs (Table 1) (Ward, 2014). 80% of T-LBLs are diagnosed at IV stage. Enlargement of mediastinal mass is the first syndrome in 60%-80% of T-LBLs. T-LBL affects bone marrow in 25% of cases, affects neck LNs in 32% of cases, and affects center nerve system in 10% of cases. B-LBL affects rarely mediastinum and does not develop into leukemia. T-LBL originates from T-lymphoblast or T-pro-lymphocyte. Having similar pathology and disease progression, T-LBL and T-cell acute lymphoblastic leukemia (T-ALL) are now classified into the same entity (You, 2015). T-LBL is sensitive to chemotherapy; and 80%-90% of pediatric T-LBLs and 45%-55% of adult cases can be cured.

So far, the causing factor for T-LBL is unknown. Over 55% of pediatric T-LBLs have chromosome changes (CCs) in tumor cells, and the most frequent CCs are *TCR*-gene-related CCs (Lones, 2006). The loci of *TCR* genes in chromosomes are: α chain in 14q11.2 and β chain in 7q34. Several forms of translocations of *TCR* genes are associated with T-LBL development. They include t(7; 9), t(10;14), t(1;14), t(5;14), and t(7;19) (Table 2). In each of these translocations, a fusion gene of *TCR* with a transcription factor is generated. The transcription factors involved in these translocations are respectively: *NOTCH1* in t(7;9), *HOX 11/TLX1* in t(10;14), *TAL1/SCL* in t(1;14), *TAL2/TLX3/HOX11L2* in t(5;14), and *LYL1* in t(7;19). As a consequence, the translocated transcription factor will be over-expressed by the promoter of *TCR* gene. Constant expression of a transcription factor can drive stimulator-independent mitosis and lead to cell transformation of a T-lymphoblast.

## 2.2 Forms of lymphomas occurring in both adults and children

HL and DLBCL are two common forms of lymphomas. They occur in adults and children (Brugières, 2016). However, HL has a peak incidence at 20s, but DLBCL does not have. DLBCL and HL occur rarely in young children (< age 5). HLs make up 20%-30% of pediatric lymphomas and DLBCLs account for 8%-10% of pediatric NHLs (Table 1). HL and DLBCL both develop from a B-cell in germinal center of a LN. However, HL is an indolent form of lymphoma but DLBCL is an aggressive one. It is thus interesting to know what a factor determines the difference between HL and DLBCL on pathology and on occurring age.

### 2.2.1 Hodgkin's lymphoma (HL)

HLs represent 10% of total lymphomas. HL has two types: classic HL and nodular lymphocyte-predominant HL (NLPHL). Classic HL and NLPHL are different by morphology and cellular characteristic of R-S cells (Tsai, 2007). The R-S cells in classic HL are named as H/R-S cells, and typical H/R-S cell is a large mirror image cell. H/R-S cells do not exhibit B-cell surface markers including CD20 (namely CD20 (-)), but they are positive for CD30 and CD5. Differently, NLPHL is characterized by popcorn-like R-S cells (named as L-H/R-S cells) (Agostinelli, 2014). L-H/R-S cells exhibit B-cell markers including CD20 (namely CD20 (+)).

DNA rearrangement of *Ig* genes is detectable in both H/R-S cells and L-H/R-S cells. 95% of HLs are classic HL. Classic HLs have four subtypes: nodular sclerotic HL (NS-HL), mixed cellular HL (MC-HL), lymphocyte-depleted HL (LD-HL), and lymphocyte-rich HL (LR-HL).

NS-HL is the most common subtype of classic HL, and 60%-70% of HLs are NS-HL. NS-HL is characterized by the deposition of connective tissues in affected LNs. NS-HL occurs mainly at age 15-35, and rarely in young children (Table 1). HL starts often from a LN and grows slowly. The most affected LNs by HL are those in the neck-supraclavicular area (75%) and those in mediastinum (in chest, 60%). About 20% of HLs have alcohol-related pain in affected LNs (Cavalli, 1998). More than 200 forms of abnormal karyotypes have been observed in HL cells; however they all have low recurrences (Jansen, 1998). H/R-S cells have often complex karyotypes (Table 2). EBV-infection is closely associated with HL development. EBV DNAs are detectable in R-S cells in over 30% of HLs (Murray, 2015). In western countries, HL and infectious mononucleosis (IM) have both high incidences in young adults, and they are both related to EBV-infection (Dunmire, 2015).

### 2.2.2 *Diffuse large B-cell lymphoma (DLBCL)*

DLBCL is one of the most aggressive forms of lymphomas in adults. DLBCLs represent 30%-40% of NHLs (Table 2). Over 40% of DLBCLs are diagnosed at advanced stage, in which the tumor has affected multiple LNs and extra-nodular LTs (Chiappella, 2016). Although DLBCL occurs also in children, most DLBCL patients are old people. DLBCL has two major subtypes: germinal center B-cell like (called GCB) and activated B-cell like (called ABC). Over 50% of adult DLBCLs are GCB and only 15% are ABC. Pediatric DLBCLs are mostly GCB. ABC has worse prognosis than GCB.

Among all DLBCLs, 30%-50% of cases have over-expression of *MYC* in tumor cells, 20%-35% have over-expression of *BCL2*, and 5%-15% have *MYC* translocation (Sarkozy, 2015). *MYC* translocation is found in 30% of pediatric GCBs but only 10% of adult GCBs (Oschlies, 2006; Gualco, 2009). A lymphoma that has two forms of chromosome translocations in tumor cells is named as "double-hit lymphoma" (DHL) (Burotto, 2016). The DHLs that have both of *BCL2* translocation and *MYC* translocation account for 62% of total DHLs (Nowakowski, 2015). Most DHLs are GCB. DHL is thought to be a form of lymphoma between BL and DLBCL. Other frequent forms of DNA changes in GCB include t(14;18) (in 30% of GCBs), *c-rel* amplification (30%), *E2H2* mutation (20%), *PTEN* mutation (10%), *TP53* mutation, and *BCL6* mutation (Dobashi, 2016). t(14;18) is found in many adult GCBs, but it is negative in pediatric GCBs. ABC does not have the above DNA changes, but 30% of ABCs have *MYD88* mutation.

### 2.3 Forms of lymphomas occurring mainly in adults

FL, MCL, MALTL, and ATLL are four forms of NHLs occurring mainly in adults. They are different by cell of origin, affected organs, and causing factor. They all originate from a mature LC: FL, MCL, and MALTL from a mature B-cell whereas ATLL from a mature T-cell. FL and MALTL are localized and indolent, but MCL and ATLL are systemic. EBV-infection is associated with developments of FL and MCL. Infection of *Helicobacter-Pylori (H. pylori)* is responsible for development of gastric MALTL. Infection of HLTV-1 is the main causing factor for ATLL development.

### 2.3.1 Follicular lymphoma (FL)

FL is an indolent form of lymphoma. FLs represent 20%-30% of total NHLs and 70% of indolent NHLs (Table 2). FL starts often from a single LN in neck and grows slowly. FL originates from a centrocyte (small cleaved B-cell) or a centroblast (small non-cleaved B-cell) in germinal center of a LN. Centroblast-originated FLs are more aggressive than centrocyte-originated. The most frequent DNA change in FL is t(14;18) (IGH/BLC2 translocation). This translocation results in increased expression of *BCL2* in FL cells (Kishimoto, 2014). However, t(14;18) is found only in centrocyte-originated FLs but not in centroblast-originated FLs. Notably, t(14;18) is also detectable in LCs of some healthy people. In normal population, the frequency of t(14;18) increases with age. However, t(14;18) is negative in young children. Probably, generation of t(14;18) in LCs is an early event of FL development (Mamessier, 2014). Many FLs have over-expression of BCL6 in tumor cells (Wagner, 2011). Other forms of gene mutations in FL cells include *CREBBP, MLL,* and *EZH2*; however these mutations have all low recurrences. If not treated, 45% of FLs will proceed into DLBCL in late stage.

### 2.3.2 Mucosa-associated lymphoid tissue lymphoma (MALTL)

MALTL is also an indolent form of lymphoma. MALTLs represent 7.5% of NHLs (Table 2). MALTL occurs mostly in stomach, but it looks different from stomach cancer. MALTL appears as multifocal and diffusing tumors in stomach wall, whereas stomach cancer grows as a single mass of tumor. In most MALTLs, the tumors are localized in stomach. Only in 30% of cases, the tumors affect neighbor LNs. But MALTL affects rarely other organs (Hu, 2016). MALTL is a form of marginal zone lymphoma, because the cell of origin of MALTL is a memory B-cell. Memory B-cells are normally localized in the marginal zone of a LN/LT or spleen. In mucosa, LTs distribute in epithelium and sub-epithelium. The LCs in mucosal LTs can be injured by *H. pylori.* Long-term infection of *H. pylori* is thought to be the main causing factor for MALTL development (Park, 2014). Recurrent DNA changes in MALTL include *IRTA1 (*immunoglobulin superfamily receptor translocation-associated 1) (in 50% of cases), t(11;18) (API2/MALT1, 30%), and t(14;18) (IgH/MALT1, 10%) (Falini, 2012; Bacon, 2007).

### 2.3.3 Mantle cell lymphoma (MCL)

MCL is an adult form of lymphoma that is aggressive and has bad prognosis. MCLs represent 6% of NHLs (Table 2). In most cases, MCL starts by enlargements of spleen and the superficial LNs. MCL is diagnosed often at advanced stage. MCL may have affected digestive mucosa, bone marrow, and peripheral blood at diagnosis. However, about 20% of MCLs appear like CLL (Martin, 2017). t(11;14) is found in 95% of MCLs, and this translocation results in generation of *IGH-CCND1* fusion gene and permanent expression of *cyclin D1* in MCL cells (Yin, 2013).

MCL has two subtypes by cell of origin: subtype I (MCL-I) and subtype II (MCL-II). MCL-I arises from a naïve B-cell in mantle zone of a follicle; and in the B-cell, *IgH* hypermutation has not occurred. MCL-II arises from a naïve B-cell in germinal center; and in the B-cell, *IgH* hypermutation has occurred. MCL-I is classic subtype of MCL, and it is aggressive (Vose, 2017). MCL-I can affect multiple LNs and extra-nodular LTs. Differently, MCL-II is indolent, thus called also leukemic non-nodal MCL. The tumor cells in MCL-II can enter blood stream, bone marrow, and spleen, leading to the occurrence of chronic leukemia. *SOX11* mutation is found only in MCL-I but not in MCL-II. Apart from t(11;14) and *SOX11* mutation, MCL cells have also other DNA changes which have low recurrences (Royo, 2011). MCL is resistant to chemotherapy (Martin, 2017). Most MCL patients can survive only 3-5 years.

*2.3.4  Adult T-cell lymphoma/leukemia (ATLL)*

ATLL is an adult form of T-cell lymphoma occurring only in certain areas of the world, including Japan, Caribbean, and South America (Table 2). ATLL has the highest incidence in Japan. Infection of human T-cell lymphotropic virus type 1 (HTLV-1) is associated with ATLL development. However, only 2%-5% of HLTV-infected individuals develop ATLL (Bangham, 2015). ATLL is diagnosed often firstly by skin lesions (in 50% of cases) and lung lesions. ATLL can affect the extra-nodal LTs in skin (including nose), blood vessels, bone marrow, and center nervous system. ATLL has four pathologic subtypes: acute (in 60% cases), lymphomatous (20%), smoldering (in skin and in lung), and chronic (Qayyum, 2014). The cell of origin of ATLL is an activated mature CD4+ T/NK-cell, such as memory Th-cell and effector Th-cell. ATLL cells have over-expression of *TAX*. So far, no specific DNA change is identified in ATLL cells. The role of HTLV-1 in ATLL development is not fully understood.

## III.    Three potential sources of cell injuries of lymphoid cells (LCs)

Repeated cell injuries and DNA injuries are the triggers for generation and accumulation of DNA changes in cells (Wang-Michelitsch, 2015). LCs can be injured at any location where they have passed. However, three sources of damage may be more specific for LCs.  These potential sources are: repeated bone-remodeling during bone-growth and bone-repair in marrow cavity, long-term thymic involution in thymus, and repeated pathogen-infections in LNs and LTs (Wang-Michelitsch, 2018a).

### 3.1    Repeated bone-remodeling during bone-growth and bone-repair in marrow cavity

Hematopoietic stem cells (HSCs) and the developing LCs in marrow are the precursors of the LCs in LNs and LTs. Thus the DNA changes generated in HSCs/LCs in marrow cavity may also contribute to lymphoma development. In our view, repeated bone-remodeling during bone-growth and bone-repair may be a source of damage to the hematopoietic cells in marrow. It is known that bones grow continuously in a child and become mature at age 22-25. Acute lymphoblastic leukemia (ALL) occurs mainly in children, and the incidence of ALL is very low after age 25 (Westergaard, 1997). This age-specific incidence of ALL suggests that bone-growth might be related to ALL development. In humans, hematopoiesis takes place mainly in marrow cavities and spongy parts of bones. Marrow cavity and spongy bone are developed and enlarged with the growth of a bone. Bone-growth is a result of repeated modeling-remodeling of bone tissues (Rockville, 2004). Bone-remodeling is a process of absorption of the bone-tissue exposed to marrow cavity by osteoclasts. Osteoclasts digest bone tissue by secreting acid substances and enzymes. Thus, it is quite possible that the substances produced during bone-remodeling perturb occasionally the hematopoietic cells in marrow cavity.

It is true that the risk of injury of a hematopoietic cell by bone-remodeling is quite low. However, the long-term (25 years) repetition of bone modeling-remodeling during bone-growth can largely increase this risk. In addition, the peak incidence of ALL is at age 2-5, and this age is also the peak age of occurrence of bone fractures in children. Young children have high incidence of bone fractures probably because they have fragile muscles/bones and frequent physical activities. Bone injuries may disturb bone-growth and increase the frequency of bone-remodeling. Thus, in a child, bone injuries (external) increase the risk of injuries of hematopoietic cells by bone-remodeling during bone-growth (internal). In an adult, bone-remodeling is promoted by bone injuries, and it may also affect hematopoietic cells.

### 3.2 Long-term thymic involution in thymus

B-cells develop in bone marrow but T-cells develop in thymus. Thus, bone-remodeling is associated more with the cell injuries of developing B-cells and HSCs. However, for the developing T-cells, long-term thymic involution may be a source of cell injuries. Thymic involution starts from age 11-12, and it proceeds continuously till adult age (Gui, 2012). Death of a great number of thymic stromal cells during thymic involution may produce occasionally toxic substances to neighboring T-cells. Although the risk of injury of a T-cell by death of stromal cells is low, long-term thymic involution can increase largely this risk. T-LBL is a form of lymphoma originated from T-lymphoblast (Ward, 2014). T-LBL has high incidence in adolescents. Mediastinal mass is the first syndrome in 60%-80% of T-LBLs. Thus, T-LBL may develop probably from thymus as a result of cell transformation of a T-lymphoblast. A trigger for transformation of a T-lymphoblast may be the cell injuries of lymphoblasts caused by thymic involution.

Taken together, repeated bone-remodeling and constant thymic involution may be two internal damaging factors for hematopoietic cells, including HSCs, developing B-cells, myeloid cells, and developing T-cells. Cell injuries by internal damage can occur also to other cells including tissue cells during body development and during inflammations. DNA changes can be also generated in some injured tissue cells. However, for a tissue cell, cell transformation occurs mainly at old age, as a co-effect of external and internal damaging factors. Differently, a LC can be transformed at young age as that seen in ALL. Thus, the effect of internal damage on a LC can be recognized.

### 3.3 Pathogen-infections in lymph nodes (LNs) and lymphoid tissues (LTs)

LNs and LTs are the organs where LCs contact pathogens. In a LN/LT, pathogen-infections should be the main cause for cell injuries of LCs. Low-concentrated LTs distribute widely in skin and mucosa. The LTs in mucosa are located in epithelium and sub-epithelium. In skin, some LTs distribute in epithelium and sub-epithelium, and some distribute around dermal veins, which are called perivascular LTs. In perivascular LTs, there are many memory and effector T-cells. The naive lymphocytes in a LT can be activated by pathogens in a similar process to that in a LN.

Three types of pathogens are known to be associated with lymphoma development: EBV, HTLV-1, and *H. pylori* (Geng, 2015, Oliveira, 2017, and Krishnan, 2014). EBV is a causing factor for B-cell-originated lymphomas including BL, HL, and DLBCL. HTLV-1 is related to ATLL development. Chronic infections of *H. pylori* are responsible for development of gastric MALTL. Compared with bacterial, viral are more carcinogenetic, because viral can proliferate in host cells and damage the host DNAs directly. However, not all the individuals that have had infections of these pathogens develop lymphoma. It may be the frequency of infections that is more critical. A proof is that immunodeficiency is a known risk factor for lymphoma development. For some forms of lymphomas, the associated pathogens may be multiple.

In different forms of lymphomas, the affected organs are different. The starting site of a lymphoma is related to the infecting route of a pathogen. For example, EBV is transmitted via saliva. Thus, EBV infects firstly mucosa of mouth then mucosa of digestive duct, airway, and the lung. This explains why BL starts from mucosa of mouth/abdomen and why HL, DLBCL, and FL start from the LNs in neck/mediastinum/abdomen. HTLV-1 is transmitted via blood, thus HTLV-1 may infect at first the perivascular T-cells in skin and in capsules of organs. *H. pylori* infect mainly stomach mucosa, thus MALTLs occur mostly in stomach.

## IV. DNA changes are generated and accumulate in cells as a consequence of repeated cell injuries

DNA changes in somatic cells are generated as consequences of cell injuries and DNA injuries. There are two major types of DNA changes: point DNA mutation (PDM, called also gene mutation) and chromosome change (CC, called also cytogentic abnormality). CC has two subtypes: structural CC (SCC) and numerical CC (NCC). Chromosome translocation, inversion, and deletion are all forms of SCCs. Gain or loss of one or more chromosomes of a cell is a form of NCC. Different types of DNA changes are generated in cells by different mechanisms. Studies showed that generation of PDM is a result of Misrepair of DNA on a double-strand DNA break (Rothkamm, 2002; Kasparek, 2011; Iliakis, 2015). Generation of SCC is a result of Misrepair of DNA on multiple DNA breaks. A NCC is generated as a consequence of dysfunction of cell division triggered by damage (Wang-Michelitsch, 2018a).

Misrepair of DNA is a result of repair of DNA, and it is essential for maintaining DNA integrity in situations of DNA injuries. A PDM/SCC is generated as a result of incorrect re-linking of a broken DNA. Namely, a PDM is generated when a broken DNA is re-linked by a wrong base-pair; and a SCC is generated when a broken DNA is re-linked by a "foreign" DNA fragment. The "foreign" DNA fragment can be a DNA segment that has fallen off from another chromosome. In our view, a PDM/SCC is made for DNA repair for cell survival, thus generation of PDM/SCC is not really a mistake. Differently, a NCC is not generated for "repair"; thus survival of a cell with a NCC is a real mistake. Unfortunately, a NCC can affect multiple genes and may cause cell transformation directly.

DNA break is the basis for generation of PDM or SCC. However, DNA rearrangement of *Ig/TCR* genes may also introduce DNA breaks in LCs. During the development and the activation of lymphocytes, two processes of DNA rearrangements are undertaken on *Ig/TCR* genes: **A.** DNA rearrangement in the genes for the variable regions of Ig/TCR in lymphoblasts; and **B.** class-switching recombination of DNA segments for *Ig-Fc* genes in B-immunoblast. In these two processes, the DNA at locus of *Ig/TCR* gene needs to be firstly cut at several points to remove the unneeded parts of DNA. DNA breaks are transiently produced at this moment. Then, DNA rearrangement will be achieved by correct re-linking of the remained segments of DNA. However, if this process is disturbed by damage, incorrect re-linking of DNA may occur. If re-linking of DNA is made by a "foreign" DNA fragment from another chromosome, a DNA translocation at the locus of *Ig/TCR* is generated.

Accumulation of DNA changes in cells is a result of repeated cell injuries and repeated cell proliferation (Wang-Michelitsch, 2015). In marrow, regeneration of HSCs and proliferation of developing LCs enables the accumulation of DNA changes in HSCs and in LCs. In a LN/LT, repeated cell injuries and repeated cell proliferation of LCs are results of repeated infections of viral or bacterial (Wang-Michelitsch, 2018a). HSCs and memory cells are stem cells regenerable for our whole lifetime. Thus, only the DNA changes that are generated or inherited in HSCs and memory cells can accumulate for a long time. However, all of the offspring cells can inherit their DNA changes. Repeated infections are often a consequence of immunodeficiency. Thus, immunodeficiency is associated with lymphoma development probably by accelerating the accumulation of DNA changes in LCs. Some studies show that the immune microenvironment in a lymph organ may contribute to the cell transformation of a LC (Fowler, 2016). It may be true that microenvironment is related to the proliferation of LCs (as an immune response); however, this proliferation is limited and non-malignant. Importantly, when a cell is transformed by DNA changes, the unlimited cell proliferation will be environment-independent.

**V.   A LC may have three pathways on cell transformation**

CCs can be classified into three groups by their effects on a LC: great-effect CCs (GECCs), mild-effect CCs (MECCs), and intermediate-effect CCs (IECCs) (Wang-Michelitsch, 2018b). GECC is a type of CC that affects one or more genes and can alone drive cell transformation. For example, the t(8;14) in BL and the t(2;5) in ALK$^+$-ALCL are probably forms of GECCs. MECC is a type of CC that affects at most one gene. Similar to a PDM, a MECC is often silent or mild for a LC. For example, the t(14;18) in FL and the t(11;18) (API2/MALT1) in MALTL are possibly forms of MECCs. PDMs and MECCs can accumulate in cells, and some of them may contribute to cell transformation. IECC is a type of CC that affects one or more genes and contributes to cell transformation. With stronger effect than a PDM/MECC, an IECC can accelerate the cell transformation driven by PDMs and MECCs. Ph translocation is a good example of IECC.

A LC may have higher survivability from DNA changes than a tissue cell. This is due to three cellular characteristics of a LC: anchor-independence for survival, inducible expression of cell surface molecules, and expression of fewer genes as being the smallest cell. The higher tolerance to DNA changes makes a LC have a risk to be transformed by a chromosome change (CC). In addition, a LC may require obtaining fewer cancerous properties for cell transformation than a tissue cell. A LC has by nature some properties similar to that of a cancer cell. These properties include anoikis-resistance, non-inhibition (of proliferation) by cell-contact, production of matrix metalloproteinase, and mobility. Thus, for transformation, a LC requires obtaining only one more property by DNA changes: the stimulator-independent mitosis. A tissue cell requires obtaining at least two properties for cell transformation: stimulator-independent mitosis and loss of cell-contact inhibition. Thus, a LC can be more rapidly transformed than a tissue cell.

Hence, a LC can be transformed not only by accumulation of PDMs and MECCs, but also possibly by GECC(s) and IECC(s). On this basis, we hypothesized in a recent paper that: a LC may have three pathways on cell transformation (Wang-Michelitsch, 2018b). The three pathways are: **a slow pathway** by accumulation of PDMs and MECCs through many generations of cells; **a rapid pathway** by a GECC in "one step" one generation of cell; and **an accelerated pathway** by accumulation of PDMs, MECCs, and IECC(s) through a few generations of cells (Box 1). This hypothesis is helpful for understanding the age-specific incidence of lymphoma and lymphoid leukemia, because the cell transformations of a LC via different pathways occur at different ages. A transformation via slow pathway occurs mainly in adults. A transformation via rapid pathway occurs at any age and has no increasing incidence with age. A transformation via accelerated pathway occurs also at any age, but it has increasing incidence with age.

**Box 1. Hypothesis: a LC may have three pathways on cell transformation**

- **Slow pathway**: by accumulation of PDMs and MECCs through many generations of cells
- **Rapid pathway**: by a GECC in "one step" in one generation of cell
- **Accelerated pathway**: by accumulation of PDMs, MECCs, and IECC(s) through a few generations of cells

PDM: point DNA mutation                    GECC: great-effect chromosome change
MECC: mild-effect chromosome change        IECC: intermediate-effect chromosome change

LCs at different developing stages may have different degrees of tolerance to DNA changes. The differentiating (immature) LCs may have higher tolerance to DNA changes, and they may have a risk to be transformed by GECC(s) or IECC(s). Thus, these cells can be transformed via all three pathways. Differentiating LCs include progenitor cells, blast cells, and pro-cytes. The non-differentiating (mature) LCs have lower tolerance to DNA changes, thus they cannot survive from a GECC and an IECC. These cells can only be transformed via slow pathway. Non-differentiating LCs include naïve lymphocytes, memory cells, and effector cells. Similarly, a tissue cell cannot survive from a GECC/IECC, thus can only be transformed via slow pathway. In slow and accelerated pathway, accumulation of PDMs, MECCs, and IECC(s) takes place mainly in regenerable HSCs and memory cells.

## VI. Age of occurrence of lymphoma is determined by the transforming pathway of a LC

Distinguishing between three pathways of cell transformation of a LC is critical for understanding pediatric lymphomas. In this part, we discuss how the age of occurrence of lymphoma is determined by the transforming pathway of a LC.

### 6.1 A lymphoma occurring mainly in adults: via slow pathway

Cell transformation of a LC via slow pathway takes place mainly at old age. FL, MCL, MALTL, and ATLL are adult forms of lymphomas. DLBCL occurs mostly in adults. They all have increasing incidences with age. Therefore, they develop mainly via slow pathway (Figure 2). The DNA changes that drive the cell transformations in these forms of lymphomas are mainly PDMs and MECCs. The final driver PDM/MECC takes place in the first transformed cell, but other driver DNA changes are mostly generated in precursor HSCs/memory cells. Thus, the cell injuries of HSCs/LCs occurred in marrow cavity and that occurred in thymus may also contribute to developments of adult lymphomas. Having a peak incidence in children or adolescents, BL, T-LBL, and ALK$^+$-ALCL may not develop via slow pathway. Having a peak incidence in young adults, HL may not develop via slow pathway. Although lymphomas occur in children, most patients of lymphomas are old people. This indicates that most forms of lymphomas develop via slow pathway as a consequence of long-term repetition of cell injuries of HSCs and LCs.

An adult form of lymphoma has often heterogeneity on tumor cells in the same patient and in different patients. Heterogeneity of cancer cells is a main reason for the poor response of adult cancers to a treatment. In our view, cancer heterogeneity is a result of random generations of secondary DNA changes in some (but not all) of the cancer cells during the slow progression of cancer. Some secondary DNA changes can make cancer cells become different to each other on cell property and on sensitivity to a treatment. In addition, in an adult cancer, the "sister cell" and some "cousin cells" of transformed cell (the first tumor cell) are a kind of pre-cancer cells, because they have the same precursor cells as this transformed one. A pre-cancer cell can be triggered to transform by chemotherapy. Existence of pre-cancer cells may a reason for cancer relapse in adult cancers. In targeted treatment of adult lymphomas, the blocking targets should be the mutations that occur earlier, since these mutations may exist in all tumor cells. However, further progression of a tumor may anyway alter the sensitivity of tumor cells to a treatment. Thus, targeted treatment may be beneficial for some patients, but the effect is limited.

## 6.2 A lymphoma occurring at any age without increasing incidence with age: via rapid pathway

Cell transformation of a LC via rapid pathway can take place at any age and has no increasing incidence with age. BL, T-LBL, ALK$^+$-ALCL, and HL are four forms of lymphomas that occur at all ages, and their incidences do not increase with age. Thus, they develop more likely via rapid pathway. Adult forms of lymphomas, including FL, MALTL, MCL, and ATLL, may not develop via this pathway. DLBCL occurs also in children, but it has increasing incidence with age. Thus, DLBCL may not develop via rapid pathway (Figure 2). Since a GECC may affect cell differentiation, rapid transformation of a differentiating (immature) LC by a GECC often leads to occurrence of an aggressive form of lymphoma, as that seen in BL, T-LBL, and ALK$^+$-ALCL.

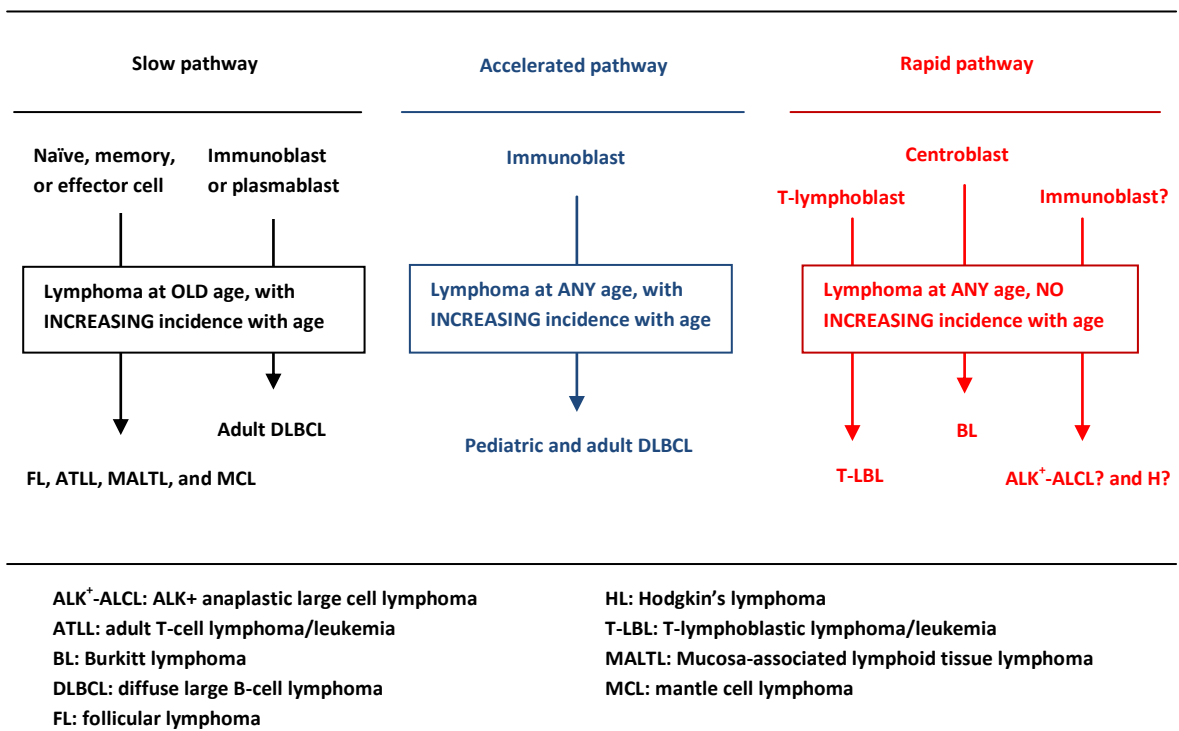

**Figure 2. The age of occurrence of lymphoma is determined by the transforming pathway of a LC**

A LC may have three pathways on cell transformation: a **rapid pathway** by a GECC, **a slow pathway** by accumulation of PDMs and MECCs, and a**n accelerated pathway** by accumulation of PDMs, MECCs, and IECC(s). Cell transformations via different pathways occur at different ages. A transformation via slow pathway takes places mainly in adults and has increasing incidence with age. FL, MCL, MALTL, and ATLL may develop via this pathway. A transformation via rapid pathway takes place at any age and has no increasing incidence with age. Rapid transformation of a blast cell results in development of an aggressive form of lymphoma, such as BL, T-LBL, and ALK$^+$-ALCL. The cells of origin in BL, T-LBL, and ALK$^+$-ALCL are respectively centroblast, T-lymphoblast, and "T-immunoblast" (?). Having a peak incidence at age 20s, HL may develop via rapid pathway from a B-immunoblast. Transformation of a LC via accelerated pathway takes place also at any age, but it has increasing incidence with age. Pediatric DLBCLs and some adult DLBCLs may develop via this pathway.

Rapid transformation of a differentiating LC may be the underlying mechanism explaining why lymphoma can occur in young children and why pediatric lymphomas are mostly aggressive. However, these pediatric forms of lymphomas, including BL, ALK$^+$-ALCL, T-LBL, and HL, have peak incidences at different ages. Their difference on age of peak incidence indicates that they may have different causing factors. A neoplasm developed via rapid pathway exhibits homogeneity of tumor cells on morphology. This homogeneity makes all tumor cells have the same sensitivity to a treatment. This explains why pediatric lymphomas have often good responses to chemotherapy.

**6.3 A lymphoma occurring at any age with increasing incidence with age: via accelerated pathway**

Transformation of a LC via accelerated pathway can take place at any age and the incidence increases with age (Figure 2). DLBCL occurs at all ages with increasing incidence with age. Thus, DLBCL can develop via this pathway. This means that DLBCL can develop via both of slow and accelerated pathway. Pediatric DLBCLs develop more likely via accelerated pathway, whereas adult DLBCLs develop via slow or accelerated pathway. In both pathways, except the final driver DNA change, most PDMs/MECCs/IECC(s) are generated in precursor cells of the first transformed cell, including precursor HSCs and memory cells. Thus, the cell injuries of HSCs/LCs occurred in marrow cavity and that of T-cells occurred in thymus may also contribute to DLBCL development. Rarely occurring in children and young people, FL, MALTL, MCL, and ATLL may not develop via this pathway. BL, T-LBL, ALK$^+$-ALCL, and HL do not have increasing incidences with age, thus they may not develop via this pathway. A cell transformation via slow or accelerated pathway begins often by low-grade cell transformation, which results in clonal proliferation. Some DLBCLs may develop from large B-cell lymphoma (LBCL), which is indolent and results from low-grade cell transformation of a B-immunoblast.

**VII. Pediatric lymphoma may develop as a result of rapid cell transformation of a LC**

To understand the age-specificity in development of lymphoma and lymphoid leukemia, we have hypothesized three pathways of a LC on transformation. In this part, we will interpret the characteristics of pediatric lymphomas by this hypothesis, in comparison with adult lymphomas. We find out that three factors may together determine the characteristics of a lymphoma, and they are cell of origin, pathway of cell transformation, and grade of cell transformation. The starting site of a lymphoma is related to the cell of origin of the lymphoma. The age of occurrence of lymphoma is determined by the pathway of cell transformation of a LC. The grade of a lymphoma is related to the grade of cell transformation of LC. A LC may have three grades of transformation: low-grade, high-grade, and intermediate-grade. A transformation is at low-grade, if cell differentiation of the transformed cell is not affected. A transformation is at high-grade, if cell differentiation is severely affected. A transformation is at intermediate-grade, if cell differentiation is partially affected.

### 7.1 BL: as a result of rapid transformation of a centroblast by severe viral infections

Endemic BL has a peak incidence at age 4-7, thus it may develop as a result of cell transformation of a LC via rapid pathway (Table 3). EBV-infection may be the main trigger for the rapid cell transformation of LC in BL development, since EBV-DNAs can be found in the tumor cells of most BLs (Rowe, 2014). The area-specific incidence of endemic BL is thought to be related to the high rate of co-infection of EBV and malaria in the children in sub-Saharan Africa (Moormann, 2016). EBV triggers cell transformation of a LC not only by inserting its DNAs into host DNAs but also by introducing DNA breaks in host cells. Multiple DNA breaks are the basis for generations of PDMs and SCCs in a LC.

Table 3. Types of driver DNA changes and pathways of cell transformation in different forms of lymphomas

| Lymphoma | Cell of origin | Driver DNA changes | Pathway of Transformation | Age of occurrence | Increasing rate with age |
|---|---|---|---|---|---|
| BL | Centroblast | A GECC, such as t(8;14) | Rapid | Any age | No |
| T-LBL | Lymphoblast/pro-lymphocyte | A GECC, such as t(10;14) | Rapid | Any age | No |
| ALK$^+$-ALCL | T-immunoblast? | A GECC, such as t(2;5) | Rapid | Any age | No |
| HL | B-immunoblast? | A GECC: complex karyotype (?) | Rapid | Any age | No |
| DLBCL | B-immunoblast | Accumulation of PDMs/MECCs/IECC(s) | Accelerated | Any age | Yes |
|  | B-immunoblast or plasmablast | Accumulation of PDMs/MECCs | Slow | > age 50 | Yes |
| FL | Centrocyte or centroblast | Accumulation of PDMs/MECCs | Slow | > age 55 | Yes |
| MCL | Naïve B- lymphocyte | Accumulation of PDMs/MECCs | Slow | > age 55 | Yes |
| MATLL | Memory B-cell | Accumulation of PDMs/MECCs | Slow | > age 55 | Yes |
| ATLL | Memory or effector Th-cell | Accumulation of PDMs/MECCs | Slow | > age 40 | Yes |

BL may arise from a centroblast, because BL cells are small non-cleaved cells. Centroblasts are proliferative and immature on cell functions, thus they may have a risk to be transformed in "one-step" by a GECC. Forms of Ig/MYC translocations, including t(8;14), t(2;8), and t(8;22), may be the main forms of GECCs that trigger BL development (Table 3). In EBV-infections, most cells that have severe DNA injuries by EBV will die, and quite a few can survive from DNA changes. Thus, transformation of a centroblast by a GECC is a rare affair, and it is paid by death of a great deal of cells. Somatic hypermutation of *Ig* genes in centroblasts may be related to the generation of Ig/MYC translocation. During hypermutation, the part of DNA containing *Ig* gene is unstable on structure and thus may be sensitive to damage (Wang-Michelitsch, 2018a).

BL starts often from mouth mucosa. There may be three reasons for that: **A.** EBV is transmitted via saliva, thus the mucosa of mouth, digestive duct, and airway wall is the first infecting area of EBV; **B.** EBV has high affinity to nasopharyngeal epithelial cells and B-cells, thus EBV can hide and proliferate in these cells in mucosa; and **C.** the LCs in mucosa may be less protected than that in LNs, thus mucosal LCs may be more frequently attacked by EBV. EBV-infection is also a causing factor for nasopharyngeal carcinoma; however, this cancer occurs mainly in adults. The difference between BL and nasopharyngeal carcinoma on age of occurrence proves that a LC has a distinct pathway on cell transformation from an epithelial cell. Namely, an epithelial cell can be transformed only via slow pathway, but a LC can be transformed not only via slow but also via rapid pathway (Wang-Michelitsch, 2018b).

### 7.2  ALK$^+$-ALCL: is it a result of rapid cell transformation of a "T-immunoblast"?

ALK$^+$-ALCL occurs mainly in children and adolescents. Having a peak incidence at age 15-20, ALK$^+$-ALCL develops more likely as a result of rapid cell transformation of a differentiating T-cell. In pathology, ALK$^+$-ALCL is characterized by cohesive proliferation of CD30+ large anaplastic blasts in affected LNs. Hence, the cell of origin of ALK$^+$-ALCL is probably a kind of "T-immunoblast", namely the precursor of effector and memory T-cells. Normally, after activation by an antigen, a naïve T-lymphocyte will firstly differentiate into T-immunoblasts in a LN/LT. A T-immunoblast will differentiate further to produce effector and memory T-cells. During the differentiation of T-immunoblasts, different generations of T-immunoblasts may have different sizes, exhibiting an anaplastic morphology of cells. A T-immunoblast is a blast cell, thus it has a risk to be transformed by a GECC such as t(2;5) (Table 3). Differently, ALK-negative ALCL and cutaneous ALCL develop mainly in adults. Thus, these two subtypes of ALCL may not develop via rapid pathway, but rather via accelerated or slow pathway.

Viral-infection may be a main causing factor for ALK$^+$-ALCL. HIV-infection is related to ALK$^+$-ALCL development in some cases. However, for most ALK$^+$-ALCLs, it is unknown which type of virus is associated. B-cell lymphomas such as BL and DLBCL affect more often mucosa, but T-cell lymphomas including ALK$^+$-ALCL and ATLL affect more often the skin. Two factors may be related to the high affection of T-cell lymphomas to skin: **A.** there are perivascular T-cells in skin and in capsules of organs; and **B.** the viral that have high affinity to T-cells may be often transmitted via blood. For example, infection of HLTV-1 is the causing factor for ATLL development. HLTV-1 is transmitted via blood. The viral of HLTV-1 can infect organs and tissues via blood circulation. Thus, the T-cells around dermal veins in skin can be repeatedly injured by HLTV-1.

## 7.3 T-LBL: is it a result of rapid transformation of a T-lymphoblast by thymic regression?

T-LBL is a rare form of lymphoma occurring mainly in older children and young adults. Having a peak incidence at age 15-20, T-LBL develops more likely as a result of rapid transformation of a T-lymphoblast. T-lymphoblasts are produced in thymus. Thymus is a small organ located in the mediastinum between two lungs. In 60%-80% of T-LBLs, the tumor starts by expansion of mediastinum mass. Thus, it is quite possible that T-LBL originates from a T-lymphoblast in thymus. As part of body development, thymus undergoes involution since age of puberty. The number of thymic stromal cells declines with age. Death of large number of stromal cells may produce toxic substances to the T-cells in thymus. Namely, the developing T-cells have a risk to be injured by thymic involution. This risk may peak at age 12-25, because the shrinking of thymus is rapid in this period of time. This explains why T-LBL develops mainly in adolescents. When T-LBL cells in thymus enter bone marrow via blood circulation, they may affect the hematopoiesis in marrow and lead to leukemia development.

The rapid transformation of T-lymphoblast in T-LBL should be driven by a GECC (Table 3). Some forms of TCR-related translocations, including t(7;9), t(7;19) t(10;14), t(1;14), and t(5;14) may be the GECCs that trigger the cell transformation of T-lymphoblast in T-LBL. Thymus carcinoma is a form of cancer that develops also in thymus but originates from an epithelial cell. Development of thymus carcinoma may be also related to thymic involution. However, thymus carcinoma occurs mainly in adults. The difference on occurring age between T-LBL and thymus carcinoma proves again that a rapid pathway of transformation of a LC exists.

## 7.4 HL: is it a result of rapid transformation of B-immunoblast by complex karyotype?

HL occurs at all ages and has a peak incidence at age 20s. Thus, HL develops more likely as a result of rapid cell transformation of a differentiating B-LC. Studies showed that H/R-S cells have a phenotype of post-germinal center B-cell, thus classic HL may originate from a B-immunoblast or a plasmablast (Rengstl, 2014; Küppers, 2002). However, as a localized form of lymphoma, HL arises more likely from a B-immunoblast. B-immunoblasts are localized in germinal center, but plasmablasts can be transported to other organs by lymph and blood circulation. Typical H/R-S cells are giant mirror cells, and other H/R-S cells have often multiple nucleus. Probably, such a giant mirror cell or multi-nucleated cell is produced by incomplete cytokinesis. Incomplete cytokinesis may be a consequence of cell injury during cell division. H/R-S cells have often non-specific forms of complex karyotypes (multiple chromosome changes). Generation of complex karyotype in a cell can be a co-effect of incomplete cytokinesis and viral-attacking to host DNAs.

Complex karyotype is a form of GECC, because it can affect multiple genes in a cell. Most forms of complex karyotypes are fatal for cells. However, for certain types of LCs which have higher survivability from DNA changes, some forms of complex karyotypes may be not fatal but rather promote cell transformation directly (Table 3 and Table 4). Hence, classic HL may develop as a result of "one-step" cell transformation of B-immunoblast by a complex karyotype. However, since complex karyotype may disturb cell function on multiple aspects, the transformed cells may proliferate slowly although independently. H/R-S cells can provoke inflammatory response by altered cell phenotypes or by produced toxic substances. Inflammatory responses can also slow down the cloning expansion of H/R-S cells. Some HL patients have alcohol-induced pain in affected LNs. Our explanation is that: by inducing vasodilatation, alcohol can accelerate infusion of lymphocytes into LNs and enhance the inflammation in HL-affected LNs.

EBV-infection is closely associated with HL development. EBV DNAs and EBV expression are detectable in H/R-S cells in 30%-50% of HLs (Geng, 2015). EBV has three effects on a LC: **A.** producing DNA breaks, **B.** inserting viral DNAs into host DNAs, and **C.** causing genome instability of host cells. For a dividing cell, EBV-attacking may disturb cell division and result in generation of complex karyotype. EBV is transmitted via saliva, thus EBV infects firstly the mucosa of mouth and airway of the victim. The LNs in neck collect the lymph from mouth and airway, thus HL starts often from a LN in the neck-supraclavicular area.

EBV-infection is the main causing factor for both of BL and HL. However, a key question is: why HL peaks at age 20s but BL peaks at age 4-7. One reason for this difference may be: the EBV-infection in BL development is severer than that in HL development. Two facts support this hypothesis: **A.** HL occurs mainly in LNs, whereas BL starts directly from mucosa; and **B.** Ig/MYC translocation has much lower opportunity to be generated in a cell than a non-specific form of complex karyotype as a consequence of DNA injuries by EBV-infection. HL has a peak incidence in young adults; however this peak incidence exists only in western countries but not in Asian countries. One explanation is that western young people may have higher rate of EBV-infections because of their habits of kissing in daily life. This can be proved by the higher incidence of EBV-associated infectious mononucleosis in western countries than in other countries. HL has also higher incidence in cities than in countryside. Our explanation is: city people may have higher frequency of EBV-infections, because of the larger population in cities and the busier social life of city people.

### 7.5 DLBCL: as a result of transformation of a B-immunoblast/plasmablast via slow or accelerated pathway

DLBCL is an aggressive form of lymphoma occurring in both children and adults. DLBCL may develop as a result of transformation of a LC via slow or accelerated pathway. DLBCL is the most common form of lymphoma possibly by two reasons: **A.** DLBCL develops via two pathways: a slow and an accelerated; and **B.** DLBCL originates from two types of cells: immunoblast in GCB and plasmablast in ABC. Immunoblasts are localized in germinal center of a LN, but plasmablasts in a LN can enter lymph and bloodstream. This can explain why GCB is less aggressive than ABC. HSCs are stem cells for immunoblasts/plasmablasts. Thus the DNA changes generated in HSCs/LCs in marrow (Table 3) may also contribute to DLBCL development.

Pediatric DLBCLs are mainly GCB. Pediatric DLBCL develops more likely via accelerated pathway, driven by accumulation of PDMs, MECCs, and IECC(s). 30% of pediatric DLBCLs have MYC-translocation, implying that MYC-translocation may be a form of IECC. Differently from the Ig/MYC translocation in BL, the MYC-translocation in DLBCL is not in locus of *Ig* gene, thus this MYC-translocation has an effect of IECC but not of GECC. An immunoblast/plasmablast may be unable to survive from Ig/MYC translocation (Table 4).

B-cell lymphomas including BL, HL, DLBCL, and FL represent about 70% of lymphomas. They are all related to EBV-infection. In EBV-infections, the viral firstly attack the lymphocytes in mucosa and then those in LNs in deep tissues. This explains why EBV-related lymphomas affect the LNs in neck, mediastinum, and abdomen. However, a key question is: EBV can attack all the LCs in a LN/LT, but why some individuals develop BL and others develop HL, DLBCL, or FL. In our view, the form of lymphoma is probably determined by two random factors: the types of LCs randomly injured by EBV-attacking and the forms of DNA changes randomly generated in LCs.

Firstly, the diversity of DNA changes generated in LCs by EBV-attacking is related to the severity of an infection. Namely, the severer an infection is, the larger diversity of DNA changes can be generated. Secondly, LCs at different differentiating stages have different tolerances to a DNA change. To the same DNA change, different LCs may have different reactions. For example, in a severe EBV-infection, t(8;14) (Ig/MYC) may be generated in some LCs in a LN/LT; however, only a centroblast may be able to survive and be transformed by t(8;14), which results in BL development (Table 4). Among all the LCs in a LN/LT, possibly only an immunoblast can survive and be transformed by a complex karyotype, which results in HL development. An immunoblast is tolerant to an IECC, thus DLBCL can develop via accelerated pathway and occur in a child. However, all LCs can survive and be transformed by long-term accumulation of PDMs and MECCs. This is the reason why adult forms of lymphomas have much higher incidences than pediatric forms of lymphomas.

Table 4. LCs at different differentiating stages have different tolerances to a DNA change

| LCs at different differentiating stages | DNA changes | | | |
| --- | --- | --- | --- | --- |
| | Accumulation of PDMs/MECCs | Accumulation of PDMs/MECCs/IECC(s) | IgH/MYC translocation (GECC) | Complex karyotype (GECC) |
| Centroblast | **FL** | ? | **BL** | ---- Cell death |
| Centrocyte | **FL** | ---- Cell death | ---- Cell death | ---- Cell death |
| B-immunoblast | **GCB-DLBCL** | **GCB-DLBCL** | ---- Cell death | **HL?** |
| Plasmablast | **ABC-DLBCL** | ? | ---- Cell death | ---- Cell death |

## 7.6 FL, MCL, MATLTL, and ATLL: related to not only the cell injuries of LCs in LNs/LTs but also that of hematopoietic stem cells (HSCs) and LCs in marrow

Fl, MCL, MALTL, and ATLL are adult forms of NHLs, and they all develop as a result of transformation of a non-differentiating (mature) LC via slow pathway (Table 3). However, they have different cells of origin: centrocyte for FL, naïve lymphocyte for MCL, memory B-cell for MALTL, and effector/memory Th-cell for ATLL. HSCs are stem cells for all the LCs in LNs/LTs. Thus, the DNA changes generated in HSCs/LCs in marrow and that in developing T-cells in thymus may also contribute to development of an adult lymphoma.

### 7.6.1 FL

FL may begin by follicular neoplasia as a result of low-grade transformation of a centrocyte. t(14;18) may contribute to the cell transformation of centrocyte in FL by increasing *BCL2* expression (Table 3). However, t(14;18) is not necessarily generated in the first transformed cell but rather in a precursor cell. t(14;18) is more possibly generated in a precursor HSC, because some healthy individuals have also t(14;18) in LCs. About 30% of FLs end up by DLBCL development. There can be two pathways on the progression of FL into DLBCL: **A.** as a result of further transformation of a FL cell; or **B.** as a result of transformation of a "normal" immunoblast/plasmablast in the FL-affected LN. Probably, progression of FL into DLBCL occurs more often via the second pathway. FL originates from a centrocyte or centroblast. Immunoblasts and plasmablasts are downstream cells of a centrocyte/centroblast (not of the transformed centrocyte). Having the same precursor LCs, an immunoblast/plasmablast may have most of the DNA changes that a centrocyte has. Thus, in a FL-affected LN, some immunoblasts/plasmablasts are a kind of "pre-cancer cells". If one of these cells is transformed by obtaining an additional driver DNA change, DLBCL may occur. In this case, DLBCL development is related to FL, but not transformed from FL.

### 7.6.2 MALTL

MALTL may develop as a result of transformation of a memory B-cell (Table 3). Long-term infections of *H. pylori* may be the main trigger for MALTL development. In MALTL patients, the tumors affect mainly stomach but not other organs. The localized development of MALTL may be related to the homing tendency of memory B-cells. It is known that new memory cells produced in a LN/LT will be transported to all organs via lymph and bloodstream. However, after lymphocyte-recirculation, most memory T/B-cells tend to home back to the locations where they are activated by an antigen (Gregor, 2017; Roy, 2002). In this way, memory cells can be quickly reactivated by the same type of antigens. However, these memory cells can be as well repeatedly injured by pathogens. For example, the memory B-cells activated by *H. pylori* may cluster to stomach mucosa. These memory cells can be reactivated by *H. pylori*, but some cells can be injured by the bacterial. By causing repeated proliferation of memory cells and repeated cell injuries, chronic infections of *H. pylori* drive the accumulation of DNA changes in memory B-cells.

### 7.6.3 MCL

MCL has two subtypes: MCL-I and MCL-II. MCL-I arises from an *IgH*-mutated naïve B-cell; whereas MCL-II originates from an *IgH*-non-mutated naïve B-cell. MCL-II is indolent and leukemia-involved, whereas MCL-I is aggressive but not leukemia-involved. These two subtypes have distinct pathologies possibly because their cells of origin have distinct properties. It is probable that the *IgH*-mutated B-cells are localized in germinal center but the *IgH*-non-mutated B-cells are not localized but rather can enter lymph and bloodstream. In addition, the indolent MCL-II may develop as a result of low-grade cell transformation of a naïve B-cell. The aggressive MCL-I may develop as a result of intermediate-grade cell transformation. t(11;14) is found in both subtypes of MCL, suggesting that t(11;14) is an early event of both MCL-I and MCL-II. Thus, t(11;14) is only associated with a low-grade cell transformation of naïve B-cell. This means that t(11;14) is more likely a form of MECC, but not a IECC. Differently, *SOX11* mutation is found only in MCL-I, suggesting that *SOX11* mutation may be related to the intermediate-grade cell transformation of B-cell in MCL-I.

DLBCL, FL, MALTL, and MCL are all B-cell lymphomas. It is known that B-cells are produced in bone marrow whereas T-cells are produced in thymus. Thus, repeated bone-remodeling during bone-growth and bone-repair may also contribute to the developments of adult forms of B-cell lymphomas. Males have higher risk of bone injuries than females due to their heavier weight and heavier physical work. This may be one reason why B-cell lymphomas occur more often in male than in female.

### 7.6.4 ATLL

Occurring mainly in adults, ATLL is more likely a result of cell transformation of a mature Th-cell by accumulation of PDMs and MECCs (Table 3). Chronic infections of HTLV-1 may be the main causing factor for generation and accumulation of PDMs and MECCs in T-cells (Ohshima, 2015). However, the DNA changes generated in precursor cells of T-cells including the HSCs in marrow and the developing T-cells in thymus may also contribute to ATLL development. ATLL has four pathologic subtypes: acute, lymphomatous, smoldering, and chronic. Different subtypes of ATLL may be results of transformation at different grades. The cell transformation in acute subtype is possibly at high-grade or intermediate-grade, that in lymphomatous subtype may be at intermediate-grade, and that in smoldering and chronic subtypes may be at low-grade. However, it is possible that ATLL begins by a smoldering or chronic subtype, and acute ATLL occurs as a result of progression of chronic ATLL.

ATLL affects often skin, and this may be related to the infecting route of HTLV-1. HTLV-1 is transmitted via blood, thus the viral can infect organs via blood circulation. In skin and capsules of organs, some T-cells distribute around small dermal veins (Ono, 2015; Nomura, 2014). HTLV-1 has high affinity to T-cells. The perivascular T-cells can be activated by the HTLV-1 in bloodstream and produce memory T-cells. By homing tendency, the memory T-cells recognizing HTLV-1 will cluster again to the perivascular areas in dermis. Thus, these memory T-cells can be repeatedly activated by HTLV-1 to produce new memory cells (Gregor, 2017; Mueller, 2013). However, these memory cells can be also repeatedly injured by HTLV-1. Repeated cell injuries and repeated cell proliferation drives the accumulation of DNA changes in memory T-cells. ATLL may occur when one of the memory cells is transformed. When ATLL cells enter bloodstream and spread to other organs and to other parts of skin, acute ATLL may occur. The skin lesions in ATLL may be generated as a consequence of the endothelial injuries made by tumor cells and the subsequent shortage of blood supply to skin.

## VIII. Conclusions

We have discussed in this paper the developing mechanisms of pediatric lymphomas in comparison with adult lymphomas. We suggest that, some pediatric lymphomas, including BL, T-LBL, ALK$^+$-ALCL, and HL, may develop as a result of "one-step" cell transformation driven by a great-effect chromosome change. Adult forms of lymphomas, including FL, MALTL, MCL, and ATLL, may develop as a result of cell transformation via slow pathway. DLBCL may develop via accelerated or slow pathway. BL, T-LBL, ALK$^+$-ALCL, and HL have all a peak incidence at young age, implying that they are each triggered by a single factor. Pathogen-attacking may be the main trigger for the cell transformations in BL, ALK$^+$-ALCL, and HL. Thymic involution may be a causing factor for T-LBL development. Adult lymphomas may develop as a co-effect of multiple-factors.

There are still phenomena in lymphoma development that are "mysterious". We cannot yet satisfactorily answer some questions, such as: why the occurrence of endemic BL is area-specific, why HL peaks at age 20s whereas BL peaks at age 4-7, and why HL is indolent. To answer these questions, further clinical investigations and experimental researches need to be undertaken.

It is a tragedy when lymphoma occurs in a child. Our analysis shows that: we have a risk of lymphoma development at young ages, because our LCs may have a risk to be transformed in "one-step". Due to some cellular characteristics, a LC may have higher survivability from DNA changes and require obtaining fewer cancerous properties for cell transformation than a tissue cell. Thus, a LC can be transformed more rapidly than a tissue cell, even in "one-step". To reduce the risk of lymphoma development, we have three advices: avoiding of immunodeficiency, avoiding of violent sports, and cryo-preserving of the HSCs in umbilical cord at birth.

## References

1. Agostinelli C, Pileri S. Pathobiology of hodgkin lymphoma. Mediterr J Hematol Infect Dis. 2014 Jun 5; 6(1):e2014040. doi: 10.4084/MJHID.2014.040. eCollection 2014. Review.PMID: 24959337
2. Ansell SM. (2015a) Hodgkin Lymphoma: Diagnosis and Treatment. Mayo Clin Proc. 90(11):1574-83. doi: 10.1016/j.mayocp.2015.07.005
3. Ansell SM. (2015b) Non-Hodgkin Lymphoma: Diagnosis and Treatment. Mayo Clin Proc. 90(8):1152-63. doi: 10.1016/j.mayocp.2015.04.025.
4. Bacon CM, Du MQ, Dogan A. Mucosa-associated lymphoid tissue (MALT) lymphoma: a practical guide for pathologists. J Clin Pathol. 2007 Apr; 60(4):361-72. Epub 2006 Sep 1. Review. PMID: 16950858
5. Bangham CR, Ratner L. How does HTLV-1 cause adult T-cell leukaemia/lymphoma (ATL)? Curr Opin Virol. 2015 Oct; 14:93-100. doi: 10.1016/j.coviro.2015.09.004. Epub 2015 Sep 27. Review. PMID: 26414684

6. Brugières L, Brice P. Lymphoma in Adolescents and Young Adults. Prog Tumor Res. 2016; 43:101-14. doi: 10.1159/000447080. Epub 2016 Sep 5. Review. PMID: 27595360
7. Burotto M, Berkovits A, Dunleavy K. Double hit lymphoma: from biology to therapeutic implications. Expert Rev Hematol. 2016 Jul; 9(7):669-78. doi: 10.1080/17474086.2016.1182858. Epub 2016 May 21. Review. PMID: 27166590
8. Cavalli F. Rare syndromes in Hodgkin's disease. Ann Oncol. 1998; 9 Suppl 5:S109-13. Review. PMID: 9926248
9. Chiappella A, Castellino A, Vitolo U. State-of-the-art Therapy for Advanced-stage Diffuse Large B-cell Lymphoma. Hematol Oncol Clin North Am. 2016 Dec; 30(6):1147-1162. doi: 10.1016/j.hoc.2016.07.002. Review. PMID: 27888872
10. Chung EM, Pavio M. Pediatric Extranodal Lymphoma. Radiol Clin North Am. 2016 Jul; 54(4):727-46. doi: 10.1016/j.rcl.2016.03.004. Review. PMID: 27265605
11. Dobashi A. Molecular Pathogenesis of Diffuse Large B-Cell Lymphoma. J Clin Exp Hematop. 2016; 56(2):71-78. Review. PMID: 27980305
12. Dunmire SK, Hogquist KA, Balfour HH. Infectious Mononucleosis. Curr Top Microbiol Immunol. 2015; 390(Pt 1):211-40. doi: 10.1007/978-3-319-22822-8_9. Review. PMID: 26424648
13. Falini B, Agostinelli C, Bigerna B, Pucciarini A, Pacini R, Tabarrini A, Falcinelli F, Piccioli M, Paulli M, Gambacorta M, Ponzoni M, Tiacci E, Ascani S, Martelli MP, Dalla Favera R, Stein H, Pileri SA. IRTA1 is selectively expressed in nodal and extranodal marginal zone lymphomas. Histopathology. 2012 Nov; 61(5):930-41. doi: 10.1111/j.1365-2559.2012.04289.x. Epub 2012 Jun 20. PMID: 22716304
14. Fowler NH, Cheah CY, Gascoyne RD, Gribben J, Neelapu SS, Ghia P, Bollard C, Ansell S, Curran M, Wilson WH, O'Brien S, Grant C, Little R, Zenz T, Nastoupil LJ, Dunleavy K. Role of the tumor microenvironment in mature B-cell lymphoid malignancies. Haematologica. 2016 May; 101(5):531-40. doi: 10.3324/haematol.2015.139493. Review. PMID: 27132279
15. Geng L, Wang X. Epstein-Barr Virus-associated lymphoproliferative disorders: experimental and clinical developments. Int J Clin Exp Med. 2015 Sep 15;8(9):14656-71. Review. PMID: 26628948
16. Gregor CE, Foeng J, Comerford I, McColl SR. Chemokine-Driven $CD4^+$ T Cell Homing: New Concepts and Recent Advances. Adv Immunol. 2017;135:119-181. doi: 10.1016/bs.ai.2017.03.001. Epub 2017 Apr 28. Review. PMID: 28826528
17. Gualco G, Weiss LM, Harrington WJ Jr, Bacchi CE. Nodal diffuse large B-cell lymphomas in children and adolescents: immunohistochemical expression patterns and c-MYC translocation in relation to clinical outcome. Am J Surg Pathol. 2009 Dec; 33(12):1815-22. doi: 10.1097/PAS.0b013e3181bb9a18. PMID: 19816150
18. Gui J, Mustachio LM, Su DM, Craig RW. Thymus Size and Age-related Thymic Involution: Early Programming, Sexual Dimorphism, Progenitors and Stroma. Aging Dis. 2012 Jun; 3(3):280-90. Epub 2012 Mar 14. PMID: 22724086
19. Hu Q, Zhang Y, Zhang X, Fu K. Gastric mucosa-associated lymphoid tissue lymphoma and Helicobacter pylori infection: a review of current diagnosis and management. Biomark Res. 2016 Jul 27; 4:15. doi: 10.1186/s40364-016-0068-1. eCollection 2016. Review. PMID: 27468353
20. Iliakis G, Murmann T, Soni A. Alternative end-joining repair pathways are the ultimate backup for abrogated classical non-homologous end-joining and homologous recombination repair: Implications for the formation of chromosome translocations. Mutat Res Genet Toxicol Environ Mutagen. 2015 Nov; 793:166-75. doi: 10.1016/j.mrgentox.2015.07.001. Review. PMID: 26520387
21. Jansen MP, Hopman AH, Haesevoets AM, Gennotte IA, Bot FJ, Arends JW, Ramaekers FC, Schouten HC.Chromosomal abnormalities in Hodgkin's disease are not restricted to Hodgkin/Reed-Sternberg cells. J Pathol. 1998 Jun;185(2):145-52. PMID 9713340
22. Kasparek TR, Humphrey TC. DNA double-strand break repair pathways, chromosomal rearrangements and cancer. Semin Cell Dev Biol. 2011 Oct; 22(8):886-97. doi: 10.1016/j.semcdb.2011.10.007. Review. PMID: 22027614
23. Kishimoto W, Nishikori M. Molecular pathogenesis of follicular lymphoma. J Clin Exp Hematop. 2014; 54(1):23-30. Review. PMID: 24942943
24. Krishnan A, Zaia JA. HIV-associated non-Hodgkin lymphoma: viral origins and therapeutic options. Hematology Am Soc Hematol Educ Program. 2014 Dec 5; 2014(1):584-9. doi: 10.1182/asheducation-2014.1.584. Review. PMID: 25696915
25. Küppers R, Schwering I, Bräuninger A, Rajewsky K, Hansmann ML. Biology of Hodgkin's lymphoma. Ann Oncol. 2002;13 Suppl 1:11-8. Review. PMID: 12078890
26. Lones MA, Heerema NA, Le Beau MM, Perkins SL, Kadin ME, Kjeldsberg CR, Sposto R, Meadows A, Siegel S, Buckley J, Finlay J, Abromowitch M, Cairo MS, Sanger WG. Complex secondary chromosome abnormalities in advanced stage anaplastic large cell lymphoma of children and adolescents: a report from CCG-E08. Cancer Genet Cytogenet. 2006 Dec; 171(2):89-96. PMID:17116485

27. Lowe EJ, Gross TG. Anaplastic large cell lymphoma in children and adolescents. Pediatr Hematol Oncol. 2013 Sep; 30(6):509-19. doi: 10.3109/08880018.2013.805347. Epub 2013 Jun 12. PMID: 23758281
28. Mamessier E, Broussais-Guillaumot F, Chetaille B, Bouabdallah R, Xerri L, Jaffe ES, Nadel B. Nature and importance of follicular lymphoma precursors. Haematologica. 2014 May; 99(5):802-10. doi: 10.3324/haematol.2013.085548. Review. PMID: 24790058
29. Martin P, Ghione P, Dreyling M. Mantle cell lymphoma - Current standards of care and future directions. Cancer Treat Rev. 2017 Jul; 58:51-60. doi: 10.1016/j.ctrv.2017.05.008. Epub 2017 Jun 13. Review. PMID: 28651117
30. Martin P, Ruan J, Leonard JP. The potential for chemotherapy-free strategies in mantle cell lymphoma. Blood. 2017 Sep 12. pii: blood-2017-05-737510. doi: 10.1182/blood-2017-05-737510. [Epub ahead of print] PMID: 28899853
31. Moormann AM, Bailey JA. Malaria - how this parasitic infection aids and abets EBV-associated Burkitt lymphomagenesis. Curr Opin Virol. 2016 Oct; 20:78-84. doi: 10.1016/j.coviro.2016.09.006. Review. PMID: 27689909
32. Mueller SN, Gebhardt T, Carbone FR, Heath WR. Memory T cell subsets, migration patterns, and tissue residence. Annu Rev Immunol. 2013; 31:137-61. doi: 10.1146/annurev-immunol-032712-095954. Epub 2012 Dec 3. Review. PMID: 23215646
33. Murray P, Bell A. Contribution of the Epstein-Barr Virus to the Pathogenesis of Hodgkin Lymphoma. Curr Top Microbiol Immunol. 2015; 390(Pt 1):287-313. doi: 10.1007/978-3-319-22822-8_12. Review. PMID: 26424651
34. Nguyen L, Papenhausen P, Shao H. The Role of c-MYC in B-Cell Lymphomas: Diagnostic and Molecular Aspects. Genes (Basel). 2017 Apr 5; 8(4). pii: E116. doi: 10.3390/genes8040116. Review. PMID: 28379189
35. Nomura T, Kabashima K, Miyachi Y. The panoply of αβT cells in the skin. J Dermatol Sci. 2014 Oct;76(1):3-9. doi: 10.1016/j.jdermsci.2014.07.010. Epub 2014 Aug 12. Review. PMID: 25190363
36. Nowakowski GS, Czuczman MS. ABC, GCB, and Double-Hit Diffuse Large B-Cell Lymphoma: Does Subtype Make a Difference in Therapy Selection? Am Soc Clin Oncol Educ Book. 2015:e449-57. doi: 10.14694/EdBook_AM.2015.35.e449. Review. PMID: 25993209
37. Ohshima K. Molecular Pathology of Adult T-Cell Leukemia/Lymphoma. Oncology. 2015; 89 Suppl 1:7-15. doi: 10.1159/000431058. Epub 2015 Nov 10. Review. PMID: 26550829
38. Oliveira PD, de Carvalho RF, Bittencourt AL. Adult T-cell leukemia/lymphoma in South and Central America and the Caribbean: systematic search and review. Int J STD AIDS. 2017 Mar; 28(3):217-228. doi: 10.1177/0956462416684461. PMID: 28178905
39. Park JB, Koo JS. Helicobacter pylori infection in gastric mucosa-associated lymphoid tissue lymphoma. World J Gastroenterol. 2014 Mar 21; 20(11):2751-9. doi: 10.3748/wjg.v20.i11.2751. Review. PMID: 24659867
40. Qayyum S, Choi JK. Adult T-cell leukemia/lymphoma. Arch Pathol Lab Med. 2014 Feb; 138(2):282-6. doi: 10.5858/arpa.2012-0379-RS. Review. PMID: 24476526
41. Ono S, Kabashima K. Novel insights into the role of immune cells in skin and inducible skin-associated lymphoid tissue (iSALT). Allergo J Int. 2015;24:170-179. Epub 2015 Sep 28. Review. PMID: 27069837
42. Rengstl B, Rieger MA, Newrzela S. On the origin of giant cells in Hodgkin lymphoma. Commun Integr Biol. 2014 Apr 3;7:e28602. doi: 10.4161/cib.28602. eCollection 2014. PMID: 25346790
43. Rochford R, Moormann AM. Burkitt's Lymphoma. Curr Top Microbiol Immunol. 2015; 390(Pt 1):267-85. doi: 10.1007/978-3-319-22822-8_11. Review. PMID: 26424650
44. Rockville (MD): Office of the Surgeon General (US); Bone Health and Osteoporosis: A Report of the Surgeon General. 2004. PMID: 20945569
45. Rothkamm K, Löbrich M. Misrepair of radiation-induced DNA double-strand breaks and its relevance for tumorIgenesis and cancer treatment (review). Int J Oncol. 2002 Aug; 21(2):433-40. Review. PMID:12118342
46. Rowe M, Fitzsimmons L, Bell AI. Epstein-Barr virus and Burkitt lymphoma. Chin J Cancer. 2014 Dec;33(12):609-19. doi: 10.5732/cjc.014.10190. Epub 2014 Nov 21. Review. PMID: 25418195
47. Royo C, Salaverria I, Hartmann EM, Rosenwald A, Campo E, Beà S. The complex landscape of genetic alterations in mantle cell lymphoma. Semin Cancer Biol. 2011 Nov; 21(5):322-34. doi: 10.1016/j.semcancer.2011.09.007. Epub 2011 Sep 18. Review. PMID: 21945515
48. Roy MP, Kim CH, Butcher EC. Cytokine control of memory B cell homing machinery. J Immunol. 2002 Aug 15;169(4):1676-82. PMID: 12165486

49. Sandlund JT. Non-Hodgkin Lymphoma in Children. Curr Hematol Malig Rep. 2015 Sep; 10(3):237-43. doi: 10.1007/s11899-015-0277-y. Review. PMID:26174528
50. Sarkozy C, Traverse-Glehen A, Coiffier B. Double-hit and double-protein-expression lymphomas: aggressive and refractory lymphomas. Lancet Oncol. 2015 Nov; 16(15):e555-67. doi: 10.1016/S1470-2045(15)00005-4. Review. PMID: 26545844
51. Shin H. Formation and function of tissue-resident memory T cells during viral infection. Curr Opin Virol. 2017 Nov 24; 28:61-67. doi: 10.1016/j.coviro.2017.11.001. [Epub ahead of print] Review. PMID: 29175730
52. Tsai HK, Mauch PM. Nodular lymphocyte-predominant hodgkin lymphoma. Semin Radiat Oncol. 2007 Jul; 17(3):184-9. Review. PMID:17591565
53. Vose JM. Mantle cell lymphoma: 2017 update on diagnosis, risk-stratification, and clinical management. Am J Hematol. 2017 Aug; 92(8):806-813. doi: 10.1002/ajh.24797. Review. PMID: 28699667
54. Wagner SD, Ahearne M, Ko Ferrigno P. The role of BCL6 in lymphomas and routes to therapy. Br J Haematol. 2011 Jan; 152(1):3-12. doi: 10.1111/j.1365-2141.2010.08420.x. Epub 2010 Nov 18. Review. PMID: 21083654
55. Wang-Michelitsch, Jicun; Michelitsch, Thomas. Cell transformation in tumor-development: a result of accumulation of Misrepairs of DNA through many generations of cells. arXiv:1505.01375.2015 Bibcode:2015arXiv:1505.01375W.
56. Wang-Michelitsch, Jicun; Michelitsch, Thomas. (2018a). Three potential sources of cell injuries of lymphoid cells associated with development of lymphoid leukemia and lymphoma. arXiv:2018
57. Wang-Michelitsch, Jicun; Michelitsch, Thomas. (2018b) Three pathways of cell transformation of a lymphoid cell: a slow, a rapid, and an accelerated. arXiv:2018
58. Ward E, DeSantis C, Robbins A, Kohler B, Jemal A. Childhood and adolescent cancer statistics, 2014. CA Cancer J Clin. 2014 Mar-Apr;64(2):83-103. doi: 10.3322/caac.21219. Epub 2014 Jan 31. PMID: 24488779
59. Yin CC, Luthra R. Molecular detection of t(11;14)(q13;q32) in mantle cell lymphoma. Methods Mol Biol. 2013; 999:211-6. doi: 10.1007/978-1-62703-357-2_14. PMID: 23666700
60. You MJ, Medeiros LJ, Hsi ED. T-lymphoblastic leukemia/lymphoma. Am J Clin Pathol. 2015 Sep; 144(3):411-22. doi: 10.1309/AJCPMF03LVSBLHPJ. Review. PMID: 26276771